\documentclass[aps,twocolumn,prl,superscriptaddress,showpacs,floatfix]{revtex4}
\usepackage{epsfig, amssymb}

\begin{document}
\advance\hoffset by  -4mm

\def\cal{\mathcal}
\def\piz{\pi ^0 }
\def\pip{\pi ^+ }
\def\pim{\pi ^- }
\def\BBbar{B\overline{B}}
\def\epem{e^+ e^-}
\def\Elsum{E_{\ell ^+}+E_{\ell ^-}}
\def\Emiss{E_{\rm miss}}
\def\pmiss{\vec{p}_{\rm miss}}
\def\Mmiss{M_{\rm miss}}
\def\GeV{{\rm GeV}}
\def\GeVc{{{\rm GeV}/c}}
\def\GeVcc{{{\rm GeV}/c^2}}
\def\MeV{{\rm MeV}}
\def\MeVc{{{\rm MeV}/c}}
\def\MeVcc{{\rm MeV/}c^2}
\def\to{\rightarrow}
\def\BR{{\cal B}}

\def\gevc{GeV/$c$}
\def\mevc{MeV/$c$}
\def\gevcc{GeV/$c^2$}
\def\mevcc{MeV/$c^2$}
\def\ra{\rightarrow}
\def\to{\rightarrow}
\newcommand{\rt}{\rightarrow}
\newcommand{\cont}{e^+e^- \rt q\overline{q}}
\newcommand{\mupmum}{\mu^+ \mu^-}
\newcommand{\pippim}{\pi^+ \pi^-}
\newcommand{\mmppx}{$(\mu^+ \mu^- \pi^+ \pi^- X)$}
\newcommand{\mmumu}{$M_{\mu^+\mu^-}$}
\newcommand{\dmtot}{$M_{\mu^+\mu^-\pi^+\pi^-}-M_{\mu^+\mu^-}$}
\newcommand{\ups}{$\Upsilon$}
\newcommand{\uIs}{$\Upsilon(1S)$}
\newcommand{\uIIs}{$\Upsilon(2S)$}
\newcommand{\uIIIs}{$\Upsilon(3S)$}
\newcommand{\utts}{$\Upsilon(2S,3S)$}
\newcommand{\uIVs}{$\Upsilon(4S)$}
\newcommand{\uIsd}{$\Upsilon(1S) \rightarrow \mu^{+} \mu^{-} $}
\newcommand{\uttsd}{$\Upsilon(2S,3S)\rightarrow\Upsilon(1S) \pi^{+} \pi^{-} $}
\newcommand{\uIVsd}{$\Upsilon(4S) \rightarrow \Upsilon(1S) \pi^{+} \pi^{-} $}
\newcommand{\uIVsdt}{$\Upsilon(4S)\to \Upsilon(1S)\pi^+\pi^- \to \mu^+ \mu^-\pi^+\pi^-$}
\newcommand{\brtot}{$\mathcal{B}(\Upsilon(4S)\to \Upsilon(1S)\pi^+\pi^-)\, = \,
(1.8 \,\pm\, 0.3(\mathrm{stat.})\, \pm \,0.2(\mathrm{sys.}))\times 10^{-4}$}
\title{\Large \rm  Observation of the decay $\Upsilon(4S) \ra \Upsilon(1S) \pi^{+} \pi^{-} $}
\date{\today}
\affiliation{Budker Institute of Nuclear Physics, Novosibirsk}
\affiliation{Chiba University, Chiba}
\affiliation{Chonnam National University, Kwangju}
\affiliation{University of Cincinnati, Cincinnati, Ohio 45221}
\affiliation{The Graduate University for Advanced Studies, Hayama, Japan} 
\affiliation{University of Hawaii, Honolulu, Hawaii 96822}
\affiliation{High Energy Accelerator Research Organization (KEK), Tsukuba}
\affiliation{Hiroshima Institute of Technology, Hiroshima}
\affiliation{University of Illinois at Urbana-Champaign, Urbana, Illinois 61801}
\affiliation{Institute of High Energy Physics, Vienna}
\affiliation{Institute of High Energy Physics, Protvino}
\affiliation{Institute for Theoretical and Experimental Physics, Moscow}
\affiliation{J. Stefan Institute, Ljubljana}
\affiliation{Kanagawa University, Yokohama}
\affiliation{Korea University, Seoul}
\affiliation{Kyungpook National University, Taegu}
\affiliation{Swiss Federal Institute of Technology of Lausanne, EPFL, Lausanne}
\affiliation{University of Ljubljana, Ljubljana}
\affiliation{University of Maribor, Maribor}
\affiliation{University of Melbourne, Victoria}
\affiliation{Nagoya University, Nagoya}
\affiliation{Nara Women's University, Nara}
\affiliation{National Central University, Chung-li}
\affiliation{Department of Physics, National Taiwan University, Taipei}
\affiliation{H. Niewodniczanski Institute of Nuclear Physics, Krakow}
\affiliation{Nippon Dental University, Niigata}
\affiliation{Niigata University, Niigata}
\affiliation{University of Nova Gorica, Nova Gorica}
\affiliation{Osaka City University, Osaka}
\affiliation{Osaka University, Osaka}
\affiliation{Panjab University, Chandigarh}
\affiliation{Peking University, Beijing}
\affiliation{RIKEN BNL Research Center, Upton, New York 11973}
\affiliation{Saga University, Saga}
\affiliation{University of Science and Technology of China, Hefei}
\affiliation{Shinshu University, Nagano}
\affiliation{Sungkyunkwan University, Suwon}
\affiliation{University of Sydney, Sydney NSW}
\affiliation{Tata Institute of Fundamental Research, Bombay}
\affiliation{Toho University, Funabashi}
\affiliation{Tohoku Gakuin University, Tagajo}
\affiliation{Tohoku University, Sendai}
\affiliation{Department of Physics, University of Tokyo, Tokyo}
\affiliation{Tokyo Institute of Technology, Tokyo}
\affiliation{Tokyo Metropolitan University, Tokyo}
\affiliation{Tokyo University of Agriculture and Technology, Tokyo}
\affiliation{Virginia Polytechnic Institute and State University, Blacksburg, Virginia 24061}
\affiliation{Yonsei University, Seoul}
  \author{A.~Sokolov}\affiliation{Institute of High Energy Physics, Protvino} 
  \author{M.~Shapkin}\affiliation{Institute of High Energy Physics, Protvino} 
  \author{K.~Abe}\affiliation{High Energy Accelerator Research Organization (KEK), Tsukuba} 
  \author{K.~Abe}\affiliation{Tohoku Gakuin University, Tagajo} 
  \author{I.~Adachi}\affiliation{High Energy Accelerator Research Organization (KEK), Tsukuba} 
  \author{H.~Aihara}\affiliation{Department of Physics, University of Tokyo, Tokyo} 
  \author{D.~Anipko}\affiliation{Budker Institute of Nuclear Physics, Novosibirsk} 
  \author{T.~Aushev}\affiliation{Swiss Federal Institute of Technology of Lausanne, EPFL, Lausanne}\affiliation{Institute for Theoretical and Experimental Physics, Moscow} 
  \author{A.~M.~Bakich}\affiliation{University of Sydney, Sydney NSW} 
  \author{E.~Barberio}\affiliation{University of Melbourne, Victoria} 
% ...
\collaboration{The Belle Collaboration}

\begin{abstract}
We study transitions between {\ups} states with the emission of
charged pions using 477~fb$^{-1}$ of data collected 
with the Belle detector at the KEKB asymmetric-energy $e^+ e^-$ collider.
We select inclusive $\Upsilon(4S) \to \mu^+ \mu^- \pi^+ \pi^- X$
events (where $X$ represents anything) and observe 
a peak in the distribution of the mass difference $\Delta M$ = ({\dmtot}). 
This peak, at  $\Delta M \,= \, (1119.2\,\pm\, 0.4)$~{\mevcc}, is identified 
as a signal for the decay {\uIVsd} with a subsequent transition {\uIsd}.
The measured product branching fraction 
$\mathcal{B}(\Upsilon(4S)\to \Upsilon(1S)\pi^+\pi^-)\times
\mathcal{B}(\Upsilon(1S)\to \mu^+\mu^-)=(4.4 \pm 0.8(\mathrm{stat.})
\pm 0.6(\mathrm{sys.})) \times 10^{-6}$.
When the PDG value for $\mathcal{B}(\Upsilon(1S)\to \mu^+\mu^-)$ is used, 
this corresponds to {\brtot} and a partial decay width 
$\Gamma(\Upsilon(4S) \rightarrow \Upsilon(1S) \pi^{+} \pi^{-})
\,=\,(3.7\, \pm \,0.6(\mathrm{stat.})\, \pm \,0.7(\mathrm{sys.}))$~keV.
\end{abstract}
\maketitle

%
The bottomonium state $\Upsilon(4S)$ has a mass above the
threshold for  $B \overline{B}$ pair
production and decays mainly into $B$-meson pairs
($\mathcal{B}(\Upsilon(4S) \to B\overline{B})\,>\,96\%$~\cite{PDG}).
However, the decay modes
$\Upsilon(4S) \rightarrow \Upsilon(mS) \pi \pi$ with $m\,=\, 1, 2$ 
should exist. 
The first preliminary evidence for these 
decays was presented in Ref.~\cite{prel}, and recently the BaBar Collaboration
has published an observation of these modes~\cite{BaBar}. A similar 
decay mode of the charmonium state $\psi(3770)$ has recently been 
observed~\cite{charm}. 
In this paper we report the observation of the decay mode
 $\Upsilon(4S) \rightarrow \Upsilon(1S) \pi^+ \pi^-$ 
from the Belle experiment. 
The results reported here supersede those 
of Ref.~\cite{prel}.

We use 477~fb$^{-1}$ of data collected on the $\Upsilon(4S)$ resonance 
and in the nearby continuum to study 
$\Upsilon(4S) \rightarrow \Upsilon(1S) \pi^+ \pi^-$ decays with 
a subsequent {\uIsd} transition. 
Charged particles are reconstructed and identified 
in the Belle detector~\cite{Belle}, which 
consists of a central drift chamber (CDC),
aerogel threshold Cherenkov counters (ACC), 
time-of-flight (TOF) scintillation counters, 
an electromagnetic calorimeter (ECL),
and a $K_L$-muon detector (KLM).
We require that charged tracks be well-measured
and consistent with originating from the interaction point.
Charged particles are assigned
a likelihood  $\mathcal{L}_i$~\cite{MUID} ($i$ = $\mu$, $\pi$, $K$)
 based on the matching of hits in
 	the KLM to the track extrapolated from the CDC, and identified as
 	muons if the likelihood ratio $P_{\mu}=\mathcal{L}_{\mu}/
(\mathcal{L}_{\mu}+\mathcal{L}_{\pi}+\mathcal{L}_K) \,> \, 0.8$,
corresponding to a muon detection efficiency of approximately 91.5\% 
over the polar angle range
$20^{\circ} \, \le  \,\theta \, \le \, 155^{\circ}$  
and the momentum range 
$0.7\, \mathrm{GeV}/c \,\le \, p  \,\le \,3.0\,\mathrm{GeV}/c$ 
in the laboratory frame. Electron identification uses a similar likelihood
ratio $P_e$~\cite{EID} based on CDC, ACC, and ECL information.
Charged particles that are not identified as muons
and have a likelihood ratio $P_e\,< \,$0.05 are treated as pions.

Identification of $\gamma$'s is based on information
from the ECL. 
Calorimeter clusters not associated with reconstructed charged tracks
and with energies greater than 50~MeV 
are considered as $\gamma$ candidates. 

Candidates for $\Upsilon(4S) \rightarrow \Upsilon(1S) \pi^+ \pi^-$
decays  with the subsequent
leptonic decay $\Upsilon(1S)\to \mu^+ \mu^-$
are selected from the standard Belle hadronic event sample. 
The most important selection criteria for this event sample
are the following:
multiplicity of charged tracks in an event $N_{\rm ch}\,\ge \,3$;
the event's visible energy 
$E_{\rm vis}\, \ge\,0.2\sqrt{s}$, where $\sqrt{s}$
is  the center-of-mass (c.m.) energy;
the sum of good cluster energies in the ECL
must satisfy   $0.1 \,\le \,E_{\rm sum}/\sqrt{s}\,\le \,0.8$;
the sum of the $z$ components of each charged track's and photon's momenta 
is required to satisfy $|P_{\rm z}|\,<\,0.5\sqrt{s}$.
The variables $E_{\rm vis}$,  $E_{\rm sum}$, $P_{\rm z}$ 
are evaluated in the c.m.\ system.

To select  $\Upsilon(1S)\to \mu^+ \mu^-$ decays, 
hadronic events are required to contain a $\mu^+\mu^-$
pair with $M_{\mu^+\mu^-}\,>\,9\,\mathrm{GeV}/c^2$ and also to satisfy 
$10.5\,\mathrm{GeV}\,<\,E_{\mathrm{vis}}^{\mathrm{lab}}\,< \,12.5\,
\mathrm{GeV}$, where $E_{\mathrm{vis}}^{\mathrm{lab}}$ is the event's 
visible energy calculated in the laboratory frame. The latter 
requirement reduces background from poorly reconstructed events.
After these requirements, $1.32 \times 10^5$ events remain. 
We then require the presence of a $\pi^+ \pi^-$ pair. 
To reduce background from $\Upsilon(1S)$ production in radiative 
return processes~\cite{rad} with the subsequent conversion of the emitted 
photon into an  $e^+ e^-$ pair 
that is  misidentified as $\pi^+ \pi^-$, we impose an
additional requirement on the angle between the pion momenta
in the laboratory system:  $\mathrm{cos} \theta_{\pi\pi}\,<\,0.95$.
The number of selected events is 1084.

\begin{figure}
  \includegraphics[width=0.4\textwidth] {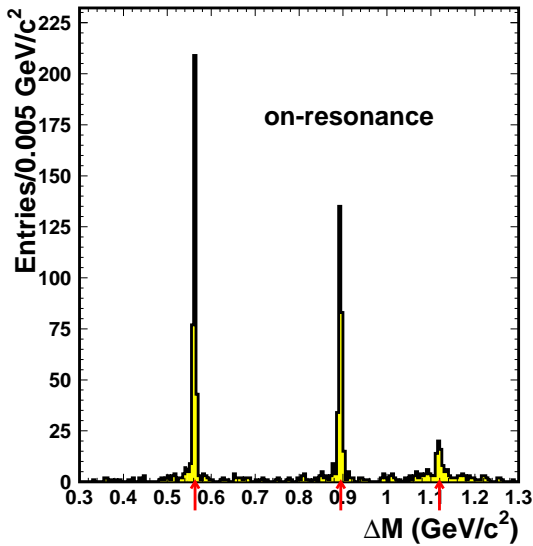}
\caption{The mass difference $\Delta M \,=\, (M_{\mu^+\mu^-\pi^+\pi^-} - 
M_{\mu^+\mu^-})$ distribution where $M_{\mu^+\mu^-}$   lies in the
$\Upsilon(1S)$ mass region.
Arrows show the positions 
of the mass differences ($m_{\Upsilon(2S)}-m_{\Upsilon(1S)}$), 
($m_{\Upsilon(3S)}-m_{\Upsilon(1S)}$), and 
($m_{\Upsilon(4S)}-m_{\Upsilon(1S)}$), respectively, based on PDG values.}
\label{fig1}
\end{figure}

To observe resonance states that decay into the  
$\Upsilon(1S) \ \pi^+ \pi^-$ final state,
the distribution of the mass difference 
$\Delta M = (M_{\mu^+\mu^-\pi^+\pi^-} - M_{\mu^+\mu^-})$ 
with $M_{\mu^+\mu^-}$ in the range
$|M_{\mu^+ \mu^-}-m_{\Upsilon(1S)}|\,< \,60\,\mathrm{MeV}/c^2$ 
is examined (see Fig.~\ref{fig1}) for the on-resonance sample.
Here $m_{\Upsilon(1S)}$ is the nominal $\Upsilon(1S)$ mass.
Three peaks are seen in the $\Delta M$ distribution. 
The first, second, and third peaks are at
$\sim$ 560, 890, and 1120 MeV/$c^2$, respectively.
The first and second peaks have very little background.
Fits to the first two peaks using Gaussians for the signal shapes
result in  peak positions of
$(561.7~\pm~0.1)\,\mathrm{MeV}/c^2$ and
$(893.4~\pm~0.2)$~MeV/$c^2$.
These values are compatible with the ($m_{\Upsilon(2S)}-m_{\Upsilon(1S)}$) and
($m_{\Upsilon(3S)}-m_{\Upsilon(1S)}$) PDG~\cite{PDG} values, respectively. 
We conclude that the first and second peaks are due to the decays 
$\Upsilon(2S) \rightarrow \Upsilon(1S) \pi^+ \pi^-$ and 
$\Upsilon(3S) \rightarrow \Upsilon(1S) \pi^+ \pi^-$, where 
the $\Upsilon(2S,3S)$ are produced mainly in the radiative return processes
$e^+e^- \rightarrow \Upsilon(2S,3S) \gamma$.  

\begin{figure}
  \includegraphics[width=0.4\textwidth] {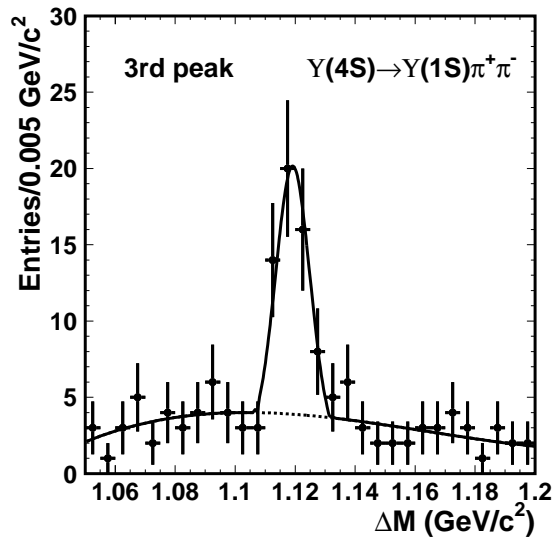}
\caption{The fit to the third peak in the $\Delta M$ distribution 
$(|M_{\mu^+ \mu^-}-m_{\Upsilon(1S)}|\,<\,60\,\mathrm{MeV}/c^2)$ 
using a Gaussian for the signal   
and a third order polynomial for the background (dotted line).
The solid line shows the sum of a Gaussian and a polynomial function.}
\label{fig2}
\end{figure}

In contrast to the first two peaks, the third peak has modest 
background.  
The position of the peak is derived from a fit to the $\Delta M$ 
distribution in the third peak region 
(see Fig.~\ref{fig2})
using a Gaussian for the signal   
and a third order polynomial for the background.
The result is $\Delta M\,= \,(1119.2 \,\pm \, 0.4)\,\mathrm{MeV}/c^2$, 
which is in good agreement 
with the mass difference  $(m_{\Upsilon(4S)}-m_{\Upsilon(1S)})$
from  the PDG.
The Gaussian width is $\sigma=(5.7\,\pm \, 1.0)\,\mathrm{MeV}/c^2$,
which is consistent with the estimated $\Delta M$ resolution.
The signal yield in the interval 
$1110\,\mathrm{MeV}/c^2\,<\,\Delta M \,< \,1135\,\mathrm{MeV}/c^2$ 
 determined
from the fit is $N_{\rm ev} \,=  \,43.9 \,\pm  \,7.9$, 
with a statistical significance of 8.0 $\sigma$ which corresponds to
${-2{\rm ln}(\mathcal{L}_0/\mathcal{L}_{\rm max})}=72.3$ 
with 3 degrees of freedom (mass, width, and yield).
Here $\mathcal{L}_0$ and
$\mathcal{L}_{\rm max}$ are the likelihood values returned by the fit with
signal yield fixed at zero and its best fit value,
respectively.
This peak is identified as a signal for the decay 
$\Upsilon(4S) \rightarrow \Upsilon(1S) \pi^+ \pi^-$
with a subsequent $ \Upsilon(1S) \rightarrow \mu^+ \mu^-$ transition. 
 
\begin{figure*}
  \includegraphics[width=0.42\textwidth] {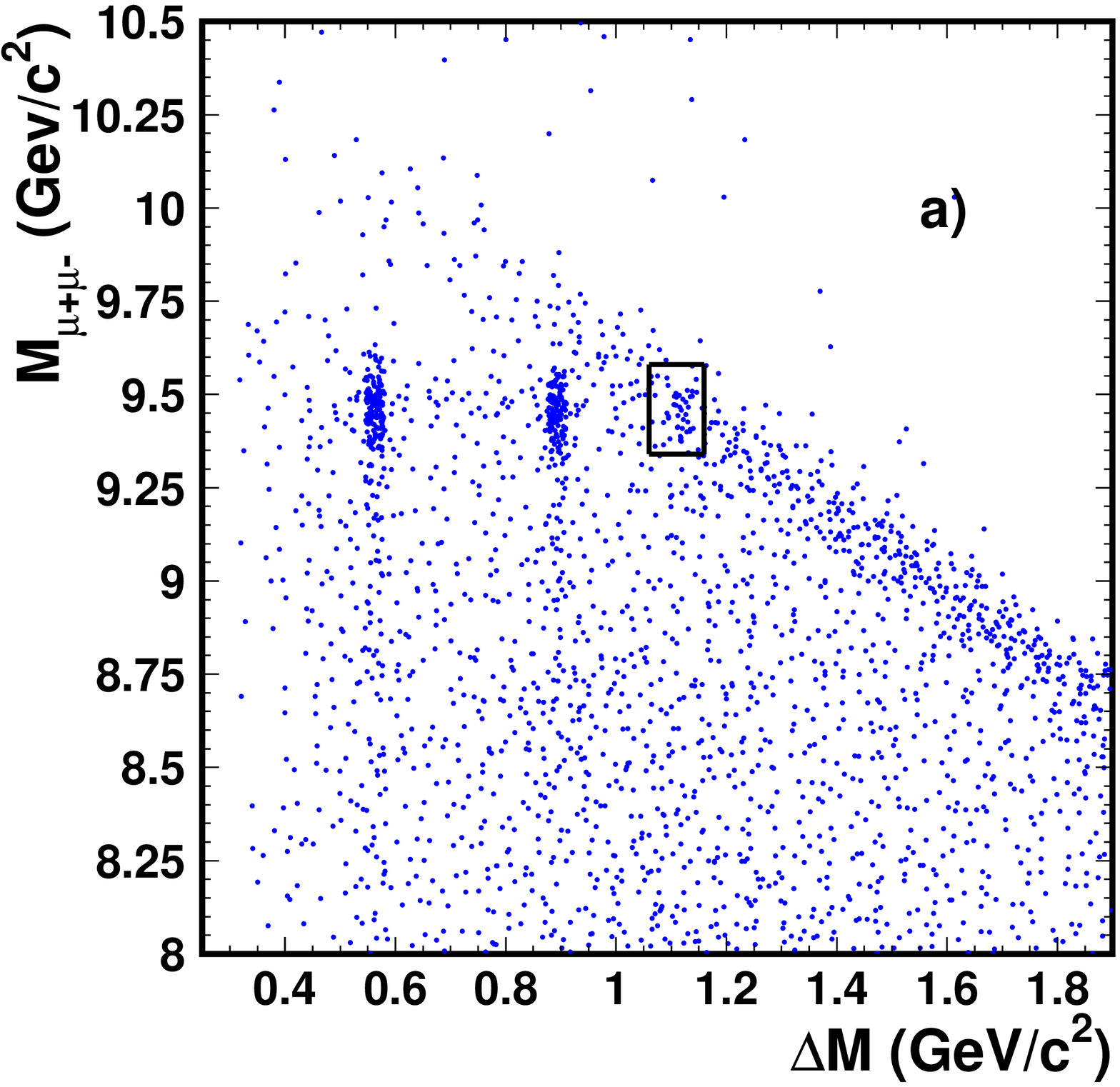} 
\hspace*{1.1cm} \includegraphics[width=0.42\textwidth] {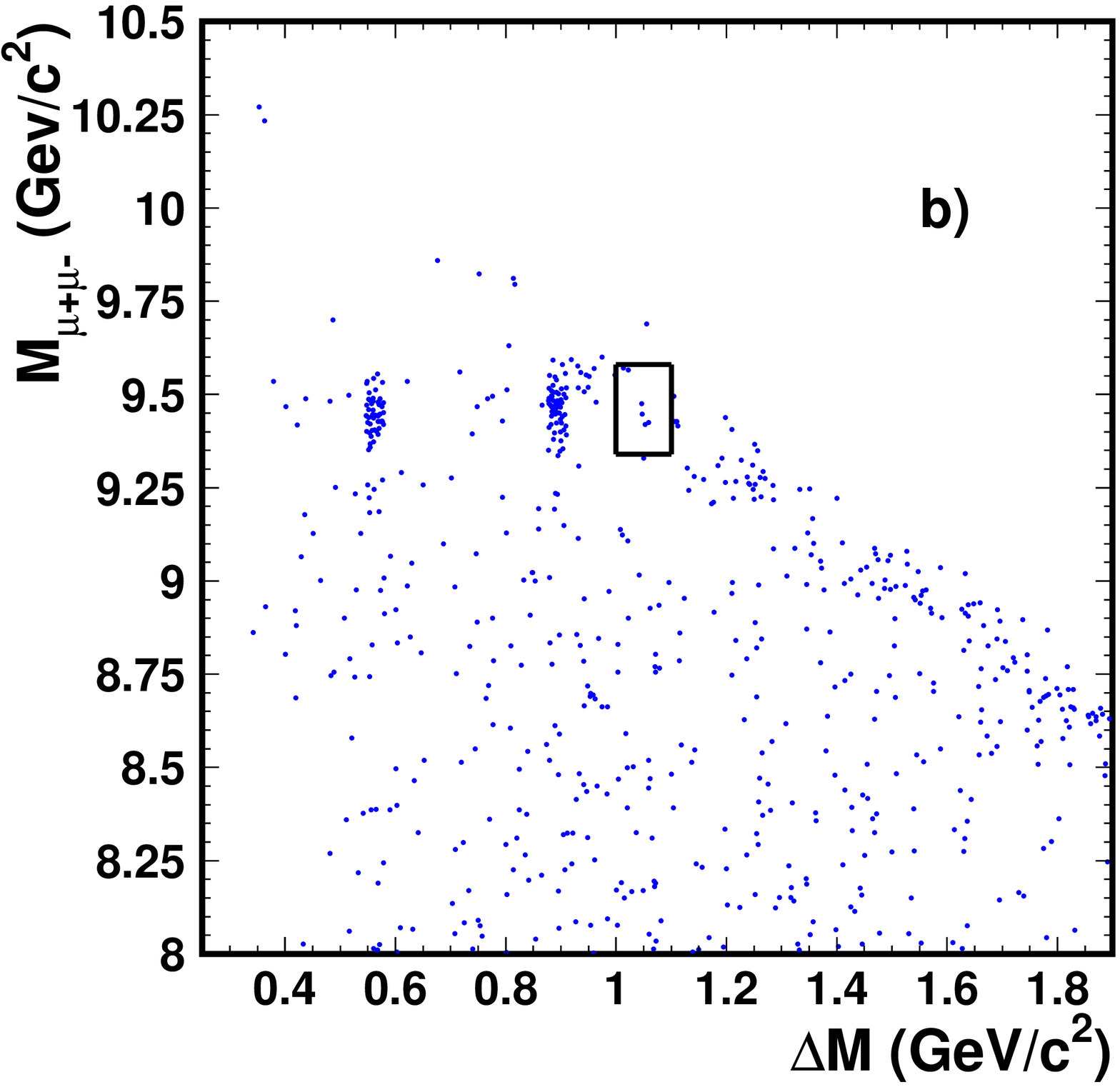}
\caption{The $\Delta M$ vs {\mmumu} scatter plot for the on-resonance 
 sample (a) and off-resonance sample (b) in the 
{\mmumu}$\,>\,8\,\mathrm{GeV}/c^2$ mass region. The rectangle in (a) 
 indicates where $\Upsilon(4S) \to \Upsilon(1S)\pi^+\pi^-$ transitions 
 would cluster; the $q\bar q$ background in this region should scale from 
 that in the rectangle in (b). For clarity, the rectangle in (a) 
 is drawn larger than the region used to determine the signal 
 yield (which is determined by optimizing the expected signal 
 yield relative to the square root of the expected
 background).
}
\label{fig3}
\end{figure*}

To verify this interpretation, 
we study the resonance properties in more detail.
First, the $\Delta M$ distribution 
for the {\mmppx} event sample
in the {\mmumu}$ \,> \,8\,\mathrm{GeV}/c^2$ mass region is considered.
Here we use a looser requirement on  {\mmumu} 
to study the background in a wider region.
The distribution of $\Delta M$ vs {\mmumu} for the on-resonance (off-resonance,
$\sqrt{s} \, =  \,10.52\,\mathrm{GeV}$, integrated luminosity
$\int \mathcal{L} dt \, = \,49.4\,\mathrm{fb}^{-1}$)
sample is shown in Fig.~\ref{fig3}a(b).  
Comparing the two distributions,
we see that the behaviour for the on-resonance and
off-resonance sample are similar except that in the off-resonance data, 
the third peak is absent.
The ``slanted band'' in this plot is due to 
the non-resonant reaction $e^+e^- \rightarrow \mu^+ \mu^- \pi^+ \pi^-$.

Using the off-resonance sample, 
the $\Delta M$ distribution
where $M_{\mu^+\mu^-}$ is restricted to 
$|M_{\mu^+ \mu^-} \,- \,m_{\Upsilon(1S)}|\,< \,60\,\mathrm{MeV}/c^2$ 
has only two peaks, corresponding to
$\Upsilon(2S)$ and  $\Upsilon(3S)$ decays.
For the on-resonance sample when the data size is scaled to the
off-resonance sample, the total number of events and the number of
background events in the interval 
$1110\,\mathrm{MeV}/c^2\,<\,\Delta M \,< \,1135\,\mathrm{MeV}/c^2$ is 
$N_{\rm tot}^{\rm res} \,= \,(7.3 \, \pm \,0.9)$ and $N_{\rm bkg}^{\rm res}
 \,= \,(2.2 \,\pm \,0.5)$, respectively. 
If we consider a $\Delta M$ interval shifted by $-60\,\mathrm{MeV}/c^2$,
 to take the lower $\sqrt{s}$ into account, we find 
$N_{\rm tot}^{\rm off} \,= \,(3.0 \,\pm \,1.7)$
events in the off-resonance sample.
To compare the scaled total
$(N_{\rm tot}^{\rm res})$ and background $(N_{\rm bkg}^{\rm res})$
yields as models for the off-resonant yield $N_{\rm tot}^{\rm off}$, 
we use the so-called likelihood  $\chi^2$~\cite{like}.
We find $\chi^2\,= \,0.26$ for $N_{\rm bkg}^{\rm res}$, and 
$\chi^2\,=\,3.26$  for  $N_{\rm tot}^{\rm res}$: 
the background interpretation of the events in the third peak region 
is therefore favoured, although the discrimination is relatively weak 
($1.7\sigma$).
This suggests that the events in the third peak region
show no sign of a similar enhancement.

\begin{figure}
  \includegraphics[width=0.4\textwidth] {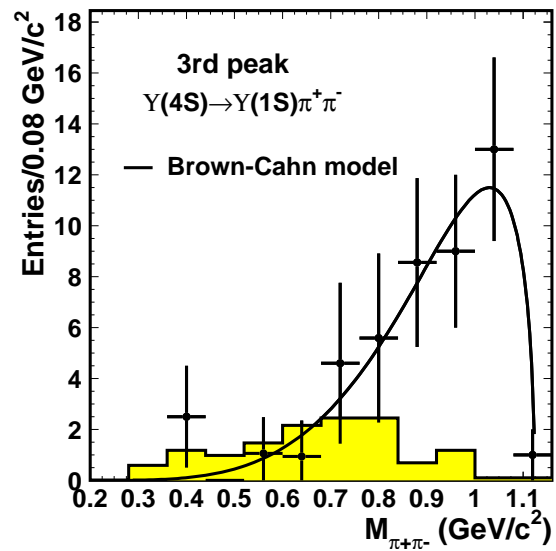}
\caption{The $\pi^+\pi^-$ invariant mass distributions
         for events from the third peak region 
   in the $\Delta M$ distribution. 
The shaded histogram is for background events estimated 
from the $\Delta M$ sideband.
The solid line shows the $\Delta M$ distribution predicted 
by the Brown-Cahn model~\cite{Brown}.}
\label{fig4}
\end{figure}

Additional information can be obtained from the study of the 
$\pi^+ \pi^-$ system.
The efficiency-corrected distribution of the invariant mass 
 of the $\pi^+\pi^-$ system  ($M_{\pi^+\pi^-}$) is shown in Fig.~\ref{fig4}
for events in the third peak region. 
The background (see Fig.~\ref{fig2})
is not subtracted from the $M_{\pi^+\pi^-}$ distribution. 
The efficiency is calculated by a Monte Carlo simulation.
The EvtGen event generator~\cite{EvtGen} with a matrix element~\cite{Brown}
taking into account particle spins is used  to produce
{\uIVsdt} events 
that are  passed through the detector
simulation and reconstruction programs.

As shown in Fig.~\ref{fig4}, the $M_{\pi^+\pi^-}$distribution for events 
from the third peak region in the $\Delta M$ distribution shows an enhancement 
    at high masses. In contrast, the distribution for background
    events, which is taken from the $\Delta M$ sideband
    ($1004\,\mathrm{MeV}/c^2\,<\,\Delta M\,<\,1110\,\mathrm{MeV}/c^2$, 
$1135\,\mathrm{MeV}/c^2\,<\,\Delta M\,<\,1210\,\mathrm{MeV}/c^2$)
 and normalized to the background events
    underneath the peak, is more uniform and shows no sign of similar
    enhancement.
This difference in the behaviour of the $M_{\pi^+\pi^-}$ distribution
is an additional argument in favour of a resonance interpretation
of the third peak.

The $M_{\pi^+\pi^-}$ distribution  can be described  
by the shape predicted
by the Brown-Cahn model~\cite{Brown} (see Fig. 4), 
where the hadronic transition between 
heavy  quarkonia is considered as a two-step process: the emission of gluons 
from heavy quarks and subsequent conversion of these gluons 
into light hadrons. 

The branching fraction for the 
$\Upsilon(4S)\to \Upsilon(1S)\pi^+\pi^-$ decay is determined from
$\mathcal{B}(\Upsilon(4S) \to \Upsilon(1S)\pi^+\pi^-)\, =\, 
N_{\rm ev}/(N_{\Upsilon(4S)}
\cdot \varepsilon \cdot \mathcal{B}(\Upsilon(1S) \rightarrow \mu^+ \mu^-)).$ 
The total number of  $\Upsilon(4S)$ in the data sample is 
$ N_{\Upsilon(4S)}\, = \,(464\, \pm\, 6) \times 10^6$, and  
the nominal branching fraction 
$\mathcal{B}(\Upsilon(1S) \rightarrow \mu^+ \mu^-)\,=\,
(2.48\,\pm\, 0.05)\%$~\cite{PDG}.
The efficiency obtained from the Monte Carlo sample is 
$\varepsilon\,=\,(2.14\,\pm\,0.06)\%$.
We apply a correction of about 8\% to $E_\mathrm{sum}/\sqrt{s}$, one of
the variables used for selecting hadronic events. The efficiency
is sensitive to this variable; the correction is designed so that
the data and MC match in this variable.

The systematic error of the reconstruction efficiency 
due to this correction is 8\%.
The systematic uncertainty in the reconstruction efficiency 
due to lack of knowledge  of the {\uIVsdt} 
decay matrix element 
is estimated by varying the parameterization of the $M_{\pi^+\pi^-}$ 
distribution. Requiring a reasonable fit to the $M_{\pi^+\pi^-}$ distribution 
we find a 3\% variation in the efficiency.
Other systematic uncertainties come from the choice of the fit range 
and the background shape (4.5\%) in the $\Delta M$ distribution, 
choice of the signal range (4\%),
choice of the $\Upsilon(1S)$ mass range (6\%),
and from the tracking efficiency (4\%).
The total systematic uncertainty 
is obtained by adding these contributions in quadrature; 
the result is 13\%.

The measured product branching fraction is \\

$\mathcal{B}(\Upsilon(4S)\to \Upsilon(1S)\pi^+\pi^-)\times
\mathcal{B}(\Upsilon(1S)\to \mu^+\mu^-)$ \\
\centerline{$=(4.4 \pm 0.8(\mathrm{stat.}) \pm 0.6(\mathrm{sys.}))
\times 10^{-6}.$} \\

The branching fraction is \\

$\mathcal{B}(\Upsilon(4S)\to \Upsilon(1S)\pi^+\pi^-)$ \\
\centerline{$=(1.8 \pm 0.3(\mathrm{stat.}) \pm 0.2(\mathrm{sys.}))
\times 10^{-4}.$} \\

\noindent We also extract the partial decay width for
$\Upsilon(4S)\to \Upsilon(1S)\pi^+\pi^-$ transition
using the world-average value of the total
width~\cite{PDG} and obtain \\

$\Gamma(\Upsilon(4S)\to \Upsilon(1S)\pi^+\pi^-)$ \\
\centerline{ $= (3.7 \pm 0.6(\mathrm{stat.}) 
\pm 0.7(\mathrm{sys.}))\: \mathrm{keV}.$} \\ 

\noindent The measured values of  
$\mathcal{B}(\Upsilon(4S)\to \Upsilon(1S)\pi^+\pi^-)$ and
$\Gamma(\Upsilon(4S)\to \Upsilon(1S)\pi^+\pi^-)$
are about twice larger than  BaBar's results~\cite{BaBar}. 

To summarize, a study of transitions between $\Upsilon$ states 
with the emission of charged pions has been performed at Belle.
The mass difference distribution
($M_{\mu^+\mu^-\pi^+\pi^-}-M_{\mu^+\mu^-}$) from the
$\mu^+ \mu^- \pi^+ \pi^- X$  event sample
for $M_{\mu^+\mu^-}$ within 
the $\Upsilon(1S)$ mass region has two peaks from
$\Upsilon(2S,3S)\to \Upsilon(1S)\pi^+\pi^-$ decays with no
background. 
A third peak at $\Delta M \,= \, (1119.2\,\pm\, 0.4)\,\mathrm{MeV}/c^2$ is
interpreted as a signal for the decay
$\Upsilon(4S)\to \Upsilon(1S)\pi^+\pi^-$
with a subsequent $ \Upsilon(1S) \rightarrow \mu^+ \mu^-$ transition. 
This final state is the first example of a non-$B \overline{B}$ decay mode 
of the $\Upsilon(4S)$.
The branching fraction 
$\mathcal{B}(\Upsilon(4S)\to \Upsilon(1S)\pi^+\pi^-)$ 
and the partial decay width  
$\Gamma(\Upsilon(4S) \rightarrow \Upsilon(1S) \pi^{+} \pi^{-})$
are measured.
The $M_{\pi^+\pi^-}$ distribution 
can be described by the shape predicted
by the Brown-Cahn model~\cite{Brown}.

We thank the KEKB group for excellent operation of the
accelerator, the KEK cryogenics group for efficient solenoid
operations, and the KEK computer group and
the NII for valuable computing and Super-SINET network
support.  We acknowledge support from MEXT and JSPS (Japan);
ARC and DEST (Australia); NSFC and KIP of CAS (contract No.~10575109 
and IHEP-U-503, China); DST (India); the BK21 program of MOEHRD, and the
CHEP SRC and BR (grant No. R01-2005-000-10089-0) programs of
KOSEF (Korea); KBN (contract No.~2P03B 01324, Poland); MIST
(Russia); ARRS (Slovenia);  SNSF (Switzerland); NSC and MOE

\end{document}